**Simulation Everywhere: An Evolutionary Expansion of Discrete-Event Modeling and Simulation research and practice**

Ikpe Justice Akpan, Ph.D.[1,*] and Godwin Etti, MBA.[2]

1. Professor of Information Systems and Business Analytics, Kent State University, 330 University Drive NE, New Philadelphia, Ohio, USA.
2. Doctoral Candidate, Faculty of Management Sciences, University of Port Harcourt, Port Harcourt, Nigeria.

* Correspondence: iakpan@kent.edu; Tel.: +1 330-672-1094.

**Abstract**: Simulation was launched in the 1950s, nicknamed a tool of "last resort." Over the years, this Operations Research (OR) method has made significant progress, and utilizing the accelerated advances in computer science (hardware and software, processing speed, and advanced information visualization capabilities) to improve simulation usability in research and practice. After overcoming the initial obstacles and the scare of outliving its usefulness in the 2000s, computer simulation has remained a popular OR tool applied in diverse industries and sectors, earning its popularity leading to the term "simulation everywhere." This study uses bibliographic data from research and practice literature to evaluate the evolutionary expansion in simulation, focusing on discrete-event simulation (DES). The results show asymmetrical but positive yearly literature out-put, broadened DES adoption in diverse fields, and sustained relevance as a scientific method for tackling old, new, and emerging issues. Also, DES is an essential tool in Industry 4.0 and plays a central role in digital transformation that has swept the industrial space, from manufacturing to healthcare and other sectors. With the emergence, ongoing adoption, and deployment of generative artificial intelligence (GenAI), future studies seek ways to integrate GenAI in DES to remain relevant and improve the modeling and simulation processes.

## 1. Introduction

Computer simulation is an operations/operational research (OR) technique often deployed to model complex, dynamic, and non-linear systems with interactions among several stochastic elements and components. Over six and a half decades ago, John Harling [1], while evaluating the launch of simulation in 1957/1958, nicknamed it a tool of "last resort" [1,2], so-called because simulation is often adopted as problem-solving or a decision support system when other OR techniques cannot produce an optimal solution [1-4]. In those circumstances, mathematical models and other OR techniques cannot model systems behavior accurately, or it is difficult to break down complex systems into simpler, manageable, analytical constructs [1,3]. Another reason for preferring the simulation method occurs when experimenting with the actual system is impossible or unsafe [3,5]. On the other hand, the complexity and costs associated with developing simulation studies can be complicated, making it a tool of 'last resort' when other OR methods fail [3,6].

Further, simulation as an analytical tool can be complex, putting the developers/experts and other stakeholders involved in the simulation and modeling project in a puzzling situation [3,5-7]. On occasions when the simulation experts have little or no background in the application domain, any 'aha moments' (moments of insight) be-come the only means of achieving breakthroughs [3,7,8]. Similarly, the potential users of simulation techniques for decision-making (e.g., managers and decision-makers) who are familiar with the application domains can also find it difficult to relate to the technical model and need to generate ideas or discover knowledge to enhance their understanding of the model and simulation [6-8].

Simulation can be classified as discrete-event or continuous, based on how the state variables change during model run or experimentation [9,10]. In a continuous simulation, variables change continuously, usually through a function in which time is a variable. Its state changes all the time, not just at the time of some discrete events [3,9]. For example, the water level in a reservoir with given inflows and outflows may change continuously. In such cases, "continuous simulation" is more appropriate, although discrete-event simulation (DES) can serve as an approximation [3,5,10]. Conversely, DES explains the state variables that change instantaneously at distinct points in time. Its events occur at intervals, and the number of events is finite [3,5,11]. Discrete simulation models address what happens to the individual elements in the system and proceed to the succeeding interacting components [3,11]. For instance, in a road traffic problem, car arrivals and departures from the traffic points occur at dis-tinct points in time, referred to as events. Nothing happens between two consecutive events, which makes the arrival and departure events discrete [3,11]. The events refer to an occurrence at a time that changes the state of the system, e.g., the arrival of a customer at the service point [12]. DES process often involves experimentation on a computer-based model that depicts a replica of a real system, with the model acting as a vehicle for experimentation, often in a trial-and-error manner [13,14]. This study focuses on the evolution and expansion of DES.

The DES growth has not been without challenges. Starting in the mid-2000s, the simulation community began to raise concerns about the future of DES practice as demonstrated in several studies (e.g., "Simulation modeling is 50, do we need a reality check?" [15], "DES, where to next?" [16], and more!). By 2010, a panel discussion entitled "DES is dead, long live agent-based simulation!" (ABS) was held during the annual OR society simulation workshop in Worchester, UK [17]. This concern was re-echoed at a panel presentation during the 2011 Winter Simulation Conference [18]. The general notion from these debates considered DES to have outlived its usefulness while welcoming the 'new bride,' ABS [17]. Although few voices rose to defend DES's continuous relevance [19], the discussion against it appeared louder till a third option arose, proposing a 'holy matrimony' between ABS and DES [20,21], or with other methods, such as systems dynamics [22,23], to combine two methods to create a hybrid model [20-23].

In the end, DES survived the obstacles and the scare of having outlived its useful-ness to remain a popular OR method used to solve diverse problems in nearly several industries. In the winter of 2017, simulation experts in academia and industry assembled at Red Rock Casino Resort & Spa, Las Vegas, USA, during the Winter Simulation Conference (WSC), entitled "WSC turns 50: Simulation everywhere" [24], pointing to the continuous expansion of simulation method. In 2020 when the coronavirus disease 2029 (COVID-19) broke out, DES became a potent OR tool in addressing diverse dis-ease infection problems, including making spread forecasts, assigning and optimizing intensive care unit beds, initiating and testing healthcare management control strategies, and making critical decisions COVID-19 vaccination problems [25-27].

This study examines the evolutionary expansion of DES as an OR technique covering the past fifteen years, the challenging period of the scare of DES extinction to the period of “simulation everywhere” through a scientometric analysis of research and practice. The study seeks to address the following four research objectives:

- RO1. Assess the evolutionary expansion of DES research.
- RO2. Analyze the research landscape to uncover the thematic structure, ap-plications, and evolution.
- RO3. Evaluate the intellectual structure to highlight the impacts of documents and sources' citations and impact.
- RO4. Analyze the social structures of DES research and identify authors' and countries' collaborations.

The rest of the paper is organized as follows: Section 2 presents DES modeling processes, activities, and tasks. Section 3 addresses the materials and methods, including the bibliometric analysis framework, data collection processes and procedures, and analysis techniques. Section 4 presents and discusses the results. The final part, Section 5, concludes the paper and identifies the potential future research areas.

## 2. Background and Overview of Discrete-Event Simulation and Modeling Processes

### 2.1. Overview of Discrete-Event Simulation

The DES processes involve designing real or imaginary operations models to analyze complex systems and behavior through experimentation. Understanding the systems' behavior by conducting experiments with a simulation model helps to evaluate strategies for improving operational outcomes and decision-making [28]. Assumptions are often made about the systems and transformed into mathematical algorithms and relationships to reveal operational functionality and performance [3,29,30]). This Section presents the modeling and simulation processes, activities, and tasks.

### 2.2. Modeling Activities, Tasks, and Processes

DES as an OR technique is used to solve complex dynamic problems where exact analytic and mathematical methods are problematic or not feasible [3,30,31]. It often involves an "experimentation on a computer-based model" of a replica system, the model acting as a "vehicle for experimentation" based on a "what if" scenario and other statistical design techniques [3,5,31]. Figure 1 presents DES modeling activities, tasks, and processes.

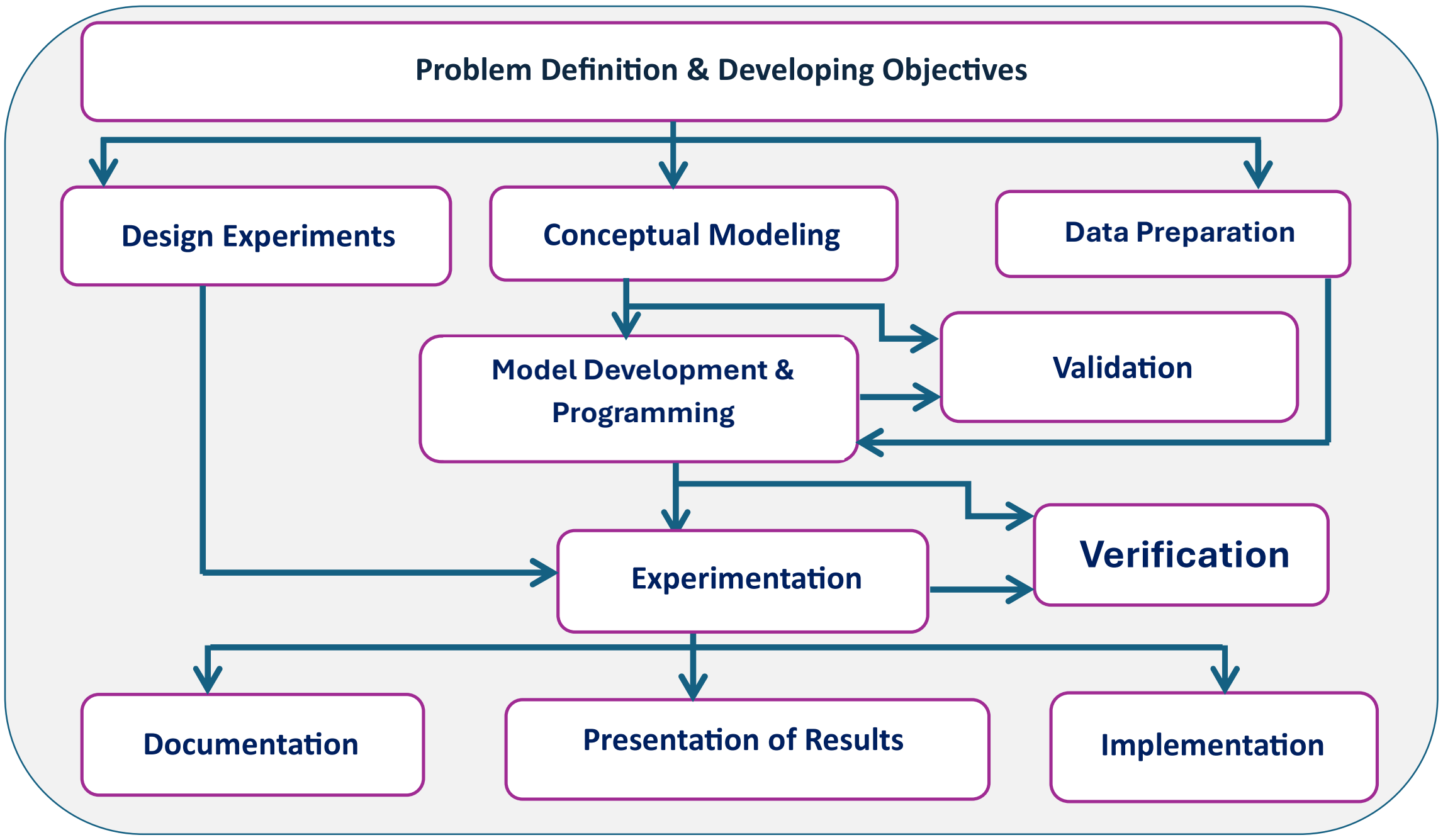

Fig. 1: Modeling and simulation tasks, activities, and processes

The first step in the simulation and modeling process involves problem definition or formulation is the first step in the simulation and modeling process. It involves analyzing the system layout and requirements, setting the objectives of the simulation study, and defining the process flow [3,5]. Other activities include preparing a list of problems to address, identifying a set of assumptions, and determining the questions the project seeks to answer based on data availability [3,30,32].

In the second step, the analyst creates a conceptual model, which is an abstract or logical representation of a proposed or imaginary operational system based on some problem specifications. The model can be represented by a diagram or pictorial sketches showing the system's process flow or layout. The analyst can also supplement the conceptual model with text or pictures [5,30,33,34]. When the conceptual modeling process employs a visual display technique, the model elements and the system's components use graphic symbols, sketches, or block diagrams supplemented by text. Further discussions about conceptual modeling activities in DES are available elsewhere [30].

Model development is the next step in the DES modeling process. This step involves implementing the conceptual model using a computer program or by 'drag and drop' of elements utilizing modeling and simulation software [3,35]. Most DES modeling and simulation efforts today use commercial software, making it possible to build models without recourse to extensive programming. The graphical images can employ 2D or 3D graphics. The parameters are also defined by using drop-down menus and command of the DES software. Examples of model-building and simulation applications today include WITNESS/VR, FlexSim, Simul8, Arena, and more [36,37].

The fourth step involves model validation. This is a process of determining whether a simulation model accurately represents the system based on set objectives [38-40]. The activities include checking and correcting errors in the model, such as logic, routing, wrong components combination, or systems errors [12,39,41]. Validation helps to ensure that the simulation model accurately mimics the real-world system. The validation process also includes designing and effecting appropriate correction mechanisms.

On the other hand, model verification is concerned with deter-mining whether the model of the system interest has implemented the conceptual model accurately together with the assumptions [34,40]. Recent studies have emphasized using 3D visualization to validate DES models. Visualizing modeled operations in 3D helps to generate insight and offers substantial help to the analyst in debugging simulation models and verifying that the model accurately represents the modeled system [39]. The visual display also enables all stakeholders to be involved in the vali-dation and verification process, as they can observe the model behavior visually and provide feedback on how well it matches the existing system [42].

Experimentation and analysis are the next steps in the DES process. It involves investigating alternative courses of action towards arriving at a preferred solution to improve the system of interest. The simulation analyst can observe the model behavior at runtime and alter parameters to examine alternatives. Other important activities include collecting output statistics, undertaking multiple replications, and undertaking statistical analysis [3,5,43]. Most experimentation and analysis also involve optimizers [5,28,43] which helps to obtain optimal solutions. Like the validation process, visualizing the model, preferably using a 3D display, can be beneficial during experimentation and analysis. Proponents posit that using 3D visualization during experimentation enhances ease of analysis and generates ideas about the modeled system. Visualizing the model at runtime and in real-time using 3D/VR during experimentation reveals the entities' actual and dynamic positions [39]. It provides true-to-scale graphics and animation, making simulation models easy to understand and invaluable for communicating new ideas and alternatives [39,41].

After experimentation comes the presentation of results and communication with clients or project owners. Visual simulation is quite effective for presenting simulation outcomes and communication with clients. Using 3D graphics and VR for these tasks can simplify the presentation and interpretation of simulation results to the users, especially to the managers and other decision-makers typically with little knowledge of statistics and computer simulation. The 3D/VR model can communicate accurate physical details, making simulation models easy to understand [36,44]. Further, using 3D graphics can simplify the presentation of the results for technical and non-technical stakeholders and decision-makers. Visualizing the model offers immense benefits in conveying ideas to senior management and the board [36]. Viewing all aspects of an operation in a 3D animated model can improve users' understanding, increasing the sense of participation and involvement of managers and other stakeholders.

### 2.3. Limitations of Simulation

DES, like other simulation methods has several limitations. Sometimes it is challenging to create simulation models that accurately replicate the actual system due to the unavailability of data to describe the system's behavior. It is common for a model to require input data that is limited or unavailable. This issue must be addressed prior to the design of the model to minimize its impact once the model is completed. Further, computer simulation can prove too expensive and time-consuming to implement. DES model is often implemented as a computer program or as some input into simulator software. Producing an efficient and workable computer program may take surprisingly long and can be complicated to implement [3]. Although general-purpose simulation software can reduce the time to produce simulation models/programs, it can be expensive to obtain.

## 3. Materials and Methods

### 3.1. Data Collection

The data used in this study came from peer-reviewed journal articles, conference proceedings, and book chapters published between the period 2010 to 2024 and indexed on the SCOPUS bibliographic database. The rationale for using the SCOPUS as the data source is that it indexes a wide coverage of literature from nearly all disciplines from quality sources [45-47]. A query string (Table 1) was created using relevant keywords on DES and applications. The use of wildcards (*) ensures that all application documents were retrieved, which were subjected to stringent filtering processes. The database survey and data collection occurred in December 2024.

The bibliographic database interface offers the option to extract as plain text (.txt), Excel (.xlsx), or comma-separated value (.csv) file. The extracted data in comma-separated file format went through further filtering, screening, and selection processes. During the screening process, non-peer-reviewed publications, duplicate entries, and irrelevant titles were removed, leaving 2077 screened publications. Table 1 shows the query string created using the search keywords and the filtering and selection criteria. Metadata was then extracted from the screened documents and exported as .csv files for processing and analysis.

Table 1. Literature survey and data collection process: Search and retrieval, filtering, screening, and selection criteria of the published documents.

| Activities/Focus | Criteria |
|---|---|
| **Data Source(s)** | SCOPUS Bibliographic Database search. |
| **Search Criteria** | (("discret*") AND ("simulat*" OR "model*")) AND PUBYEAR: 2010-2024. The search generated 187,727 published documents. |
| | **Documents Filtering, Screening, and Selection** |
| **Filtering** | Removed: Books:409, Erratum:122, Retracted:113, Letter:78, Note:71; Editorial:67; Short survey:38, Data paper:32 [187,727-930] = 186797documents.<br>Removed: Non "Discrete-Event Simulation in OR": 187,797- 184,666 = 2131 |
| **Screening** | Screened out 9 Irrelevant Documents as follows: Literature not addressing the topic of interest: 2131-54=2077 |
| **Final Documents Selection** | 2077 publications from SCOPUS published between 2010 and 2024 (during COVID-19). Documents retrieved in text formats (.txt and .csv files) for analysis. |

### 3.2. Data Analytics Tools and Techniques

Recent advances in big data analytics and the availability of software solutions (including open-source software) offer more effective and efficient ways to carry out data analysis. This study employs VOSviewer [48,49] and the BIBLIOMETRIX application, a bibliographic analysis software library embedded in the R-Studio environment [50].

The above two applications help to produce quantitative bibliometric outputs, network analysis, and visualization [48-50]. Quantitative data analytics includes statistical evaluation of publications, trends, and performances and mapping the relationships among the published documents [51,52]. Network analysis and visualization can also enrich result presentation from bibliometric analyses [53,54]. The performance measures include publication trends, citation impacts, and collaboration indexes [51]. The science mapping and evaluation offers co-citation analysis and co-occurrence of keywords analysis to identify the research streams and themes. Bibliometric coupling and co-authorship analyses highlight relationships and social interactions among publications, authors, affiliations, and countries [55,56]. These techniques help to analyze the conceptual, intellectual, and social structures of the scientific literature production on DES.

## 4. Results

### 4.1. Sample Description and Preliminary Results

Table 2 presents the summary statistics analyzed using the bibliographic metadata extracted from the 2,077 SCPs (journal articles: 885 or 42.6%; conference proceedings: 1,132 or 54.5%; book chapters: 60 or 2.9%). The results also highlight 19.3% international cooperation among co-authors across several regions, showing that DES topics attract a global audience among researchers and practitioners and affirm simulation everywhere. Also, the published documents appeared in 947 sources authored/co-authored by 5,627 researchers from 89 countries. The analysis of results in Section 4 helps answer the research questions (RO1 to RO4) presented earlier in Section 1.

**Table 2.** Descriptive statistics of the sample and preliminary results.

| Variable Description | Results | Variable Description Contd. | Results |
|---|---|---|---|
| Years of Publications | 2010-2024 | **Documents Contents:** | |
| Sources (Journals, Proceedings, Book Chapters) | 947 | Keywords Plus (ID) | 10,575 |
| **Documents Information:** | | Author's Keywords (DE) | 4,121 |
| • Articles | 885 (42.6%) | **Authors and Collaboration:** | |
| • Book chapters | 60 (2.9%) | Authors | 5,629 |
| • Conference papers | 1132 (54.5%) | Authors of single-authored docs | 36 |
| Annual publication growth rate: | 2.03% | Single-authored docs | 124 |
| Average citations per doc | 9.5 | Co-Authors per Doc | 3.62 |
| References | 50,886 | International co-authorships % | 19.31 |

### 4.2. Scientific Literature Production Trend on DES

This Section examines the first research objective (RO1) about the evolutionary expansion of DES research evidenced by scientific literature production (SLP). The analysis of SLP identifies the yearly publications covering the past fifteen years (2010 to 2024), as shown in Figure 3a. The results show an annual average SLP of 138.5. There were more conference proceedings than journal articles by about 12%. However, this is not unusual in the technology fields, where initial innovations are often presented at international conferences before appearing in Journals. While the SLP in the first three years was at or slightly below the annual average, the subsequent years recorded above-average annual SLP. The highest number of publications for any single year occurred in 2020, during the outbreak of COVID-19, when scientists utilized DES as a potent tool to analyze the various pandemic-related problems [25,26]. Overall, the analysis using R-Bibliometrics shows annual productivity trend of 2.03% (Table 2).

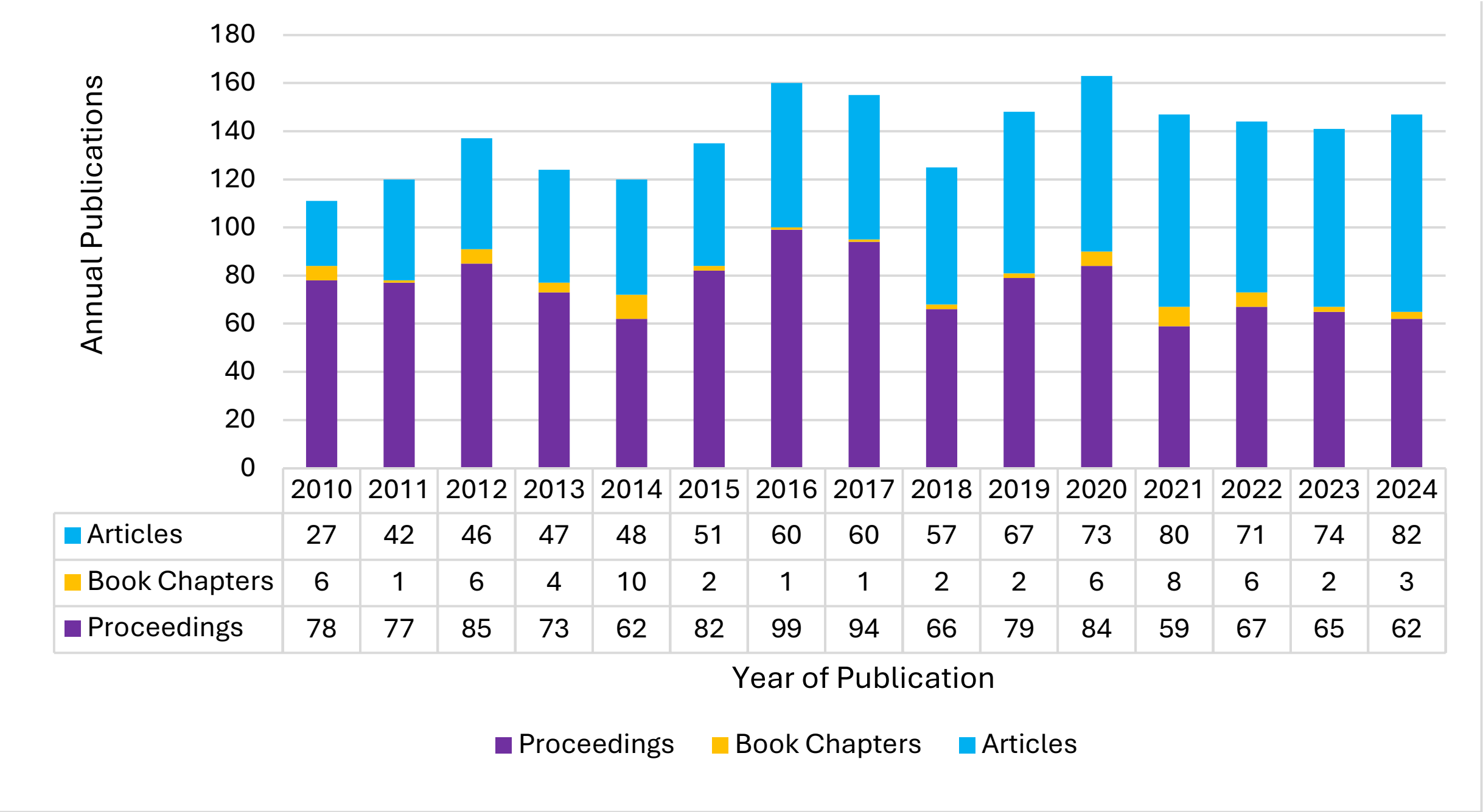

| | 2010 | 2011 | 2012 | 2013 | 2014 | 2015 | 2016 | 2017 | 2018 | 2019 | 2020 | 2021 | 2022 | 2023 | 2024 |
|---|---|---|---|---|---|---|---|---|---|---|---|---|---|---|---|
| Articles | 27 | 42 | 46 | 47 | 48 | 51 | 60 | 60 | 57 | 67 | 73 | 80 | 71 | 74 | 82 |
| Book Chapters | 6 | 1 | 6 | 4 | 10 | 2 | 1 | 1 | 2 | 2 | 6 | 8 | 6 | 2 | 3 |
| Proceedings | 78 | 77 | 85 | 73 | 62 | 82 | 99 | 94 | 66 | 79 | 84 | 59 | 67 | 65 | 62 |

Figure 2a. Annual scientific publications on discrete-event modeling and simulation.

Figure 2b presents a scatter plot of the SLP for the same period to visualize the literature productivity trend, which fits a linear trend. While the SLP for some of the years is at the mean for some years, others are slightly above or below it. The linear model produced the regression equation y = 2.232x + 120.61, where x is the nth year, and y is the total SLP. The coefficient of the determination ($R^2$ = 0.40) indicates that time predicts about 40% of the variation in SLP (annual increases). The model is appropriate for this dataset because the trend of the SLP over time exhibits growth.

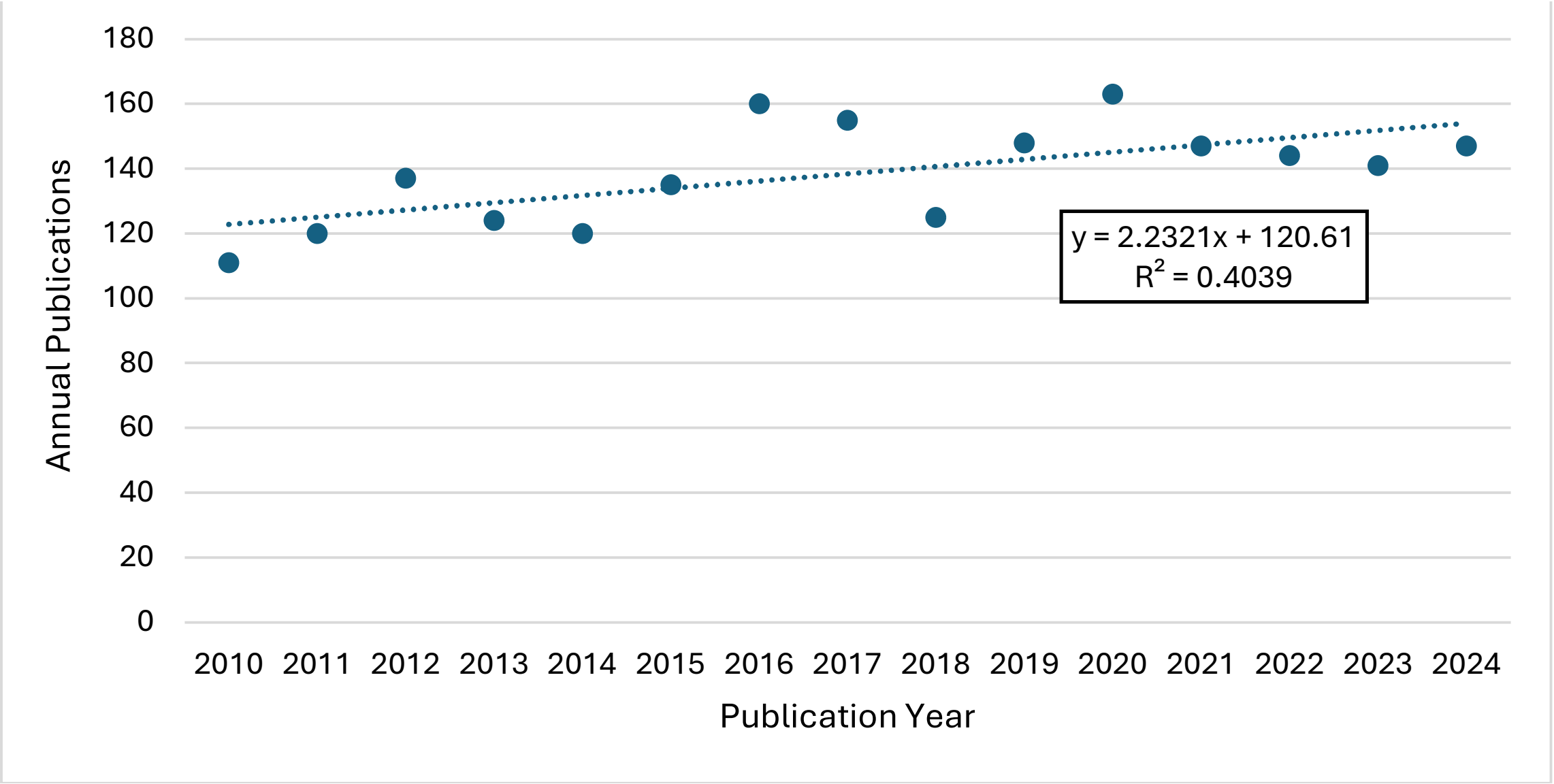

Figure 2b. Linear trend of the scientific publications on evolutionary expansion of discrete-event modeling and simulation (2010-2024).

### 4.3. Domains of Discrete-Event Simulation Research and Practice

The second research objective (RO2) analyzes the research landscape to uncover the thematic structure of DES, its applications, and evolution. This Section evaluates the DES application domains. DES as an OR technique can be utilized to solve problems in several sectors and fields of endeavor and to support decision-making in operations and production activities. The SCOPUS bibliographic data identifies enormous application areas (over 100). However, we collapsed and merged related areas (e.g., computer science theory, applications, and more) into one category named "computer science." Figure 3 presents the summary of the identified domains, some of which include computer sciences (26%), engineering (21%), mathematics, business management and decision sciences (20), medical and health sciences, computer science and engineering, business and economics, and arts and humanities. Figure 3 shows twelve disciplines where DES plays a significant role in problem-solving, especially systems optimization.

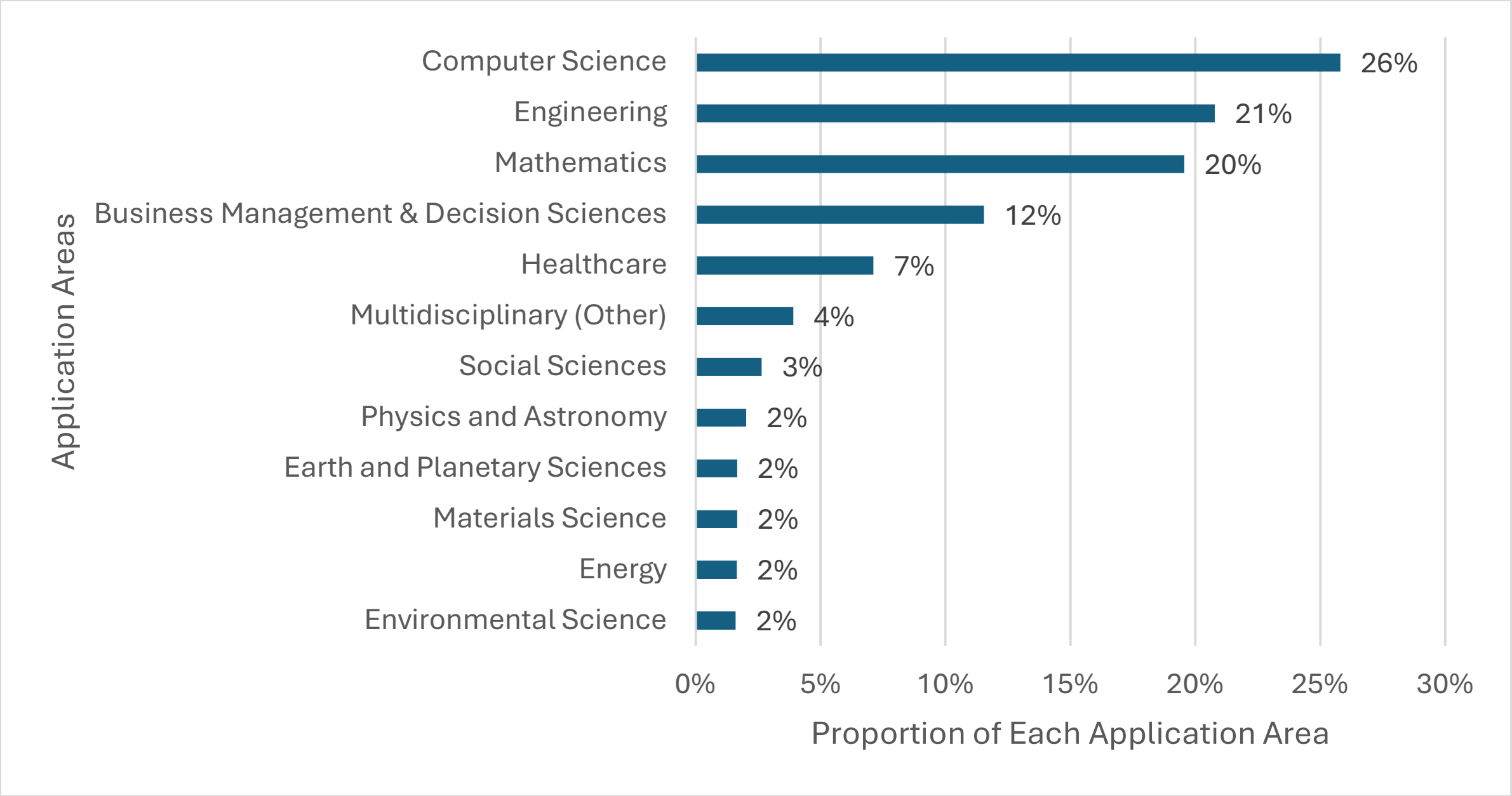

Figure 3: Discrete-event simulation application domains

**4.4 Thematic Structure of DES Research and Evolution**

This Section analyzes the thematic structure and evolutionary trends through a science mapping of DES research and practice. The preliminary results of the text analytics using the R-Bibliometrix identifies 4121 unique but unstemmed author keywords with total word frequency of 7,501 extracted from 2,077 SLPs.

The basis for using authors' keywords as an approximation to the thematic structure of research is established in the literature [50-54]. To ensure exhaustive themes analytics, the sample is stratified into three categories based on word frequency (f), which explains the number of times each keyword re-occurs in the dataset. The categorizations are as follows:

- Prominent themes, defined by prominent keywords with frequency (f) of 10 or more, ($f \geq 10$).
- Emerging themes, having keywords frequency between 4-9 ($4 \geq f < 10$).
- Least Frequent Themes ($f < 4$) and evolutionary trend during the period (2010-2024).

#### 4.4.1. Prominent Themes

The text analytics using the R-Bibliometrix highlight the most popular research themes on DES in the past fifteen years (from the period of the threat of DES extinction (discussed in the introduction) and the period of "simulation everywhere." The results identify 53 unstemmed author keywords with a total word frequency of 1,991 (f = 1991) or 26.5%. These terms, which represent the research themes on DES and modeling, went through a "stemming" process. Words 'stemming' is a text analytics process involving merging terms with different spellings but the same meaning or synonyms, e.g., "modeling/modelling" [57]. The stemming process reduced the themes from 53 to 44 (Table 3). The three (3) most popular themes other than "DES," "simulation," or "modeling" include "optimization," "healthcare," and "logistics," appearing 61, 35, and 31 times, respectively. This result shows that DES continues to tackle problems in core OR/management sciences areas.

Table 3. Analysis of eminent keywords and research themes trends in ITOR

| Keywords | Stemmed Co-Words | Link Strength | Keywords Contd. | Stemmed Co-Words | Link Strength |
|---|---|---|---|---|---|
| Discrete-Event Simulation | 973 | 494 | Waiting Time | 14 | 18 |
| Simulation | 213 | 176 | Productivity | 13 | 25 |
| Parallel Discrete-Event Simulation (DES) | 71 | 42 | Operations Research | 13 | 20 |
| Optimization | 61 | 90 | Digital Twin | 13 | 15 |
| Modeling | 55 | 83 | Performance Evaluation | 13 | 13 |
| Healthcare | 35 | 52 | Time Warp | 13 | 12 |
| Logistics | 31 | 37 | Maintenance | 12 | 22 |
| Emergency Department | 29 | 42 | Patient Flow | 12 | 18 |
| Industry 4.0 | 27 | 33 | Plant Simulation | 12 | 18 |
| Devs | 24 | 13 | Machine Learning | 12 | 14 |
| System Dynamics | 23 | 32 | Cost-Effectiveness | 12 | 11 |
| Manufacturing | 22 | 33 | Resource Allocation | 11 | 18 |
| Simulation Modeling | 20 | 22 | Supply Chain Management | 11 | 17 |
| Performance | 19 | 20 | Construction Management | 11 | 13 |
| Arena | 18 | 29 | Efficiency | 11 | 13 |
| Supply Chain | 18 | 26 | Discrete Event Systems | 11 | 6 |
| Discrete-Event | 18 | 21 | Queuing Theory | 10 | 16 |
| Covid-19 | 17 | 28 | Process Improvement | 10 | 12 |
| Scheduling | 17 | 24 | Synchronization | 10 | 11 |
| Computer Simulation | 17 | 10 | Discrete Event | 10 | 8 |
| Simulation Optimization | 15 | 11 | Process Mining | 10 | 7 |
| Design Of Experiments | 14 | 21 | Simulation Model | 10 | 7 |

Similarly, terms such as "industry 4.0," "digital twin," and "COVID-19") being among the prominent themes indicate the continuous relevance of DES in tackling problems in emerging fields and technologies in the fourth industrial revolution and disease outbreak, such as the 'stars coronavirus-2' (SARS-CoV-2) [25].

**4.4.2. Emerging themes**

This Section analyzes the emerging themes based on author keywords that occurred between 4 and 9 times (4 ≥ f < 10). The text analytics results using the R-Bibliometrix show 160 emerging themes with a total frequency of 869 times (f = 869 or 11.6% of the total word frequencies). Also, using the VOSviewer application, we created a network map and visualization of the thematic structure of DES and applications scientific literature landscape. The text analytics solution offers the functionality to remove nugatory and non-connected nodes and words that do not convey contextual meanings, leaving 148 terms. The text mining algorithm stratifies the themes into thirteen (13) color-coded clusters. The nodes with the same color belong to the same cluster throughout the network (Figure 4).

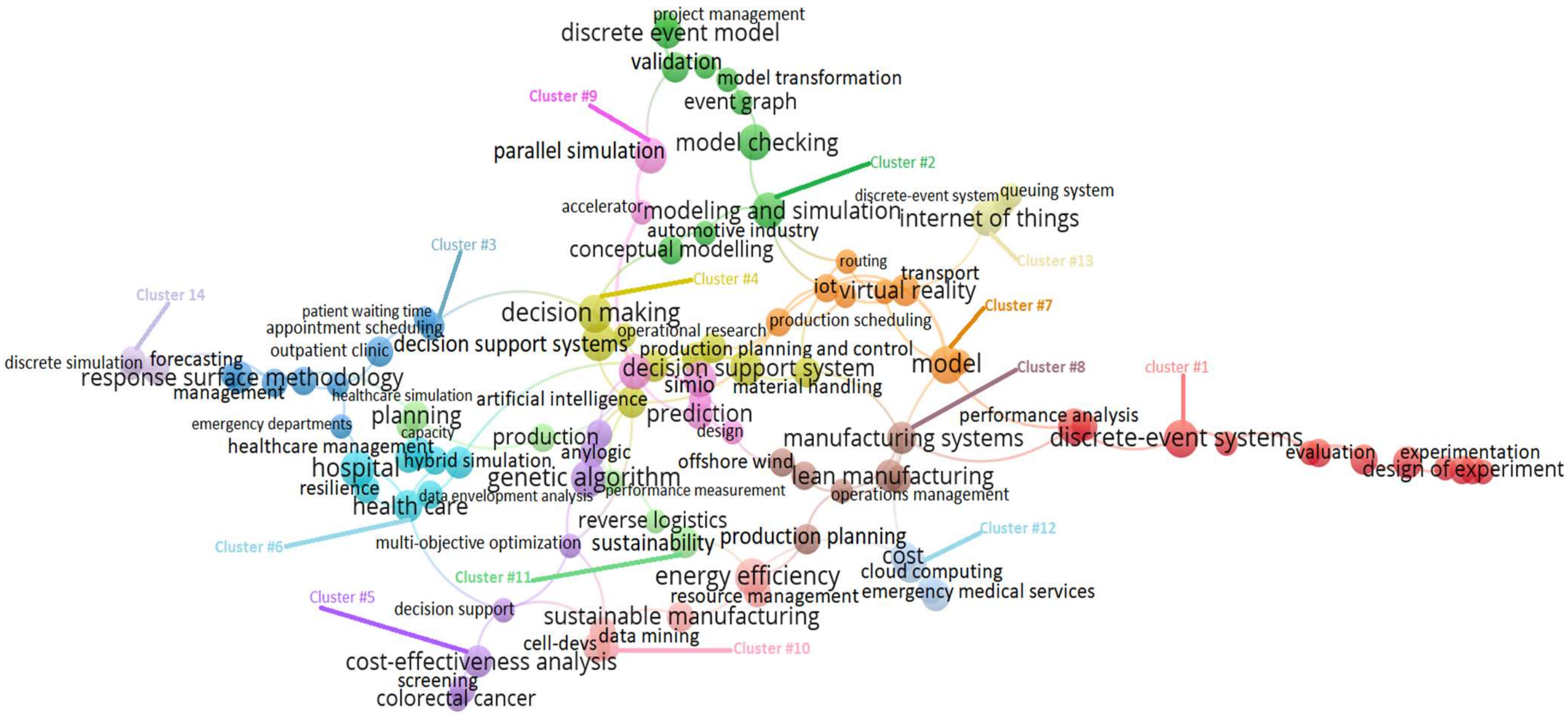

Fig. 4. Visualization of emerging research themes and co-occurrence of author keywords.

The cluster categorization contains some random elements. Removing a theme with several connected edges in the network can lead to a reclassification of the clusters. The identified research themes are more significant than the cluster where the terms are listed. The network visualization highlights the thirteen clusters as follows:

**Cluster #1**: ("discrete-event system," "design of experiment," and "experimentation." Others are "evaluation" and "performance analysis"). These terms point to conducting simulation experiments to analyze designs and systems' behavior or complex processes through what-if scenarios.

**Cluster #2:** ("modeling and simulation," "conceptual modeling," "validation," and "model checking") point to two crucial "modeling and simulation" activities: developing the conceptual model and checking errors in the programmed model, often carried out during experimentation explained above.

**Cluster #3:** The third cluster deals with "healthcare simulation" for "outpatient clinic" to determine "patient waiting time" and "appointment scheduling," a vital area of DES application.

**Cluster #4:** ("operational research," "decision support systems," "artificial intelligence," "production planning and control").

**Cluster #5:** ("cost-effectiveness analysis," "multi-objective optimization," "colorectal cancer"). These keywords point to model optimization, systems evaluation, and cancer diagnoses in healthcare.

**Cluster #6**: The cluster focuses on the use of "hybrid simulation" and "data envelopment analysis" to analyze "hospital" operations, "healthcare management," and other "healthcare" issues.

**Cluster #7**: ("production scheduling," "routing," "transport"). DES plays key roles in efficient work scheduling, vehicle routing, and product scheduling.

**Cluster #8**: ("operations management," "production planning," "lean manufacturing," and "manufacturing systems"). These themes show another aspect where DES is utilized to streamline operations.

**Cluster #9**: ("decision Support system," "prediction," "design"). The themes in this cluster relate to DES's role as a tool for aiding decision-making in facility design and making predictions.

**Cluster #10**: ("energy efficiency," "energy efficiency," "sustainable manufacturing," and "resource management").

**Cluster #11**: ("production," "reverse logistics," "sustainability") DES plays an important role as a tool to enhance sustainable and efficient production.

**Cluster #12**: ("emergency medical services," "cost," and "cloud computing"). Modeling and simulation of emergency services in healthcare and operations cost reduction forms a significant area of DES application.

**Cluster #13**: ("discrete-event system," "queuing system," and "Internet of Things"). The thirteenth cluster examines the DES application to evaluate queuing systems and the Internet of Things in Industry 4.0.

**Cluster #14**: ("discrete simulation" and "forecasting"). The two themes address the use of DES to make forecasts. Figure 4 (above) presents the social network map showing all the themes as labeled.

#### 4.4.3. Least Frequent Themes and Trends

The least frequent themes category contains terms with word frequency less than ($f < 4$). That is, the themes or keywords occurred three times or less. The dataset contains 3908 terms, making up 95% (3908/4121) of the themes. In terms of word frequencies, it represents 4641 or 62% of the 7501 total keyword frequencies The keywords are unstemmed, meaning that the text analytics algorithm considers words with similar meanings but spelled differently are considered unique (e.g., "AI" and "artificial intelligence" are considered unique terms). The text analytics algorithm categorizes the keywords into color-coded clusters on a network map. The circles (nodes) with the same color belong to the same Cluster. Also, on the legend (Figure 5), the node labels represent the research themes under focus in each period segment mapped to the period of occurrence or year of publication (pre-2014, that is, 2010 to 2014; 2014-2016; 2016-2018; 2018-2020; 2020-2024) as explained below.

The network map contains three categories of themes: DES theory, modeling and simulation processes, tasks and activities, and the diverse application areas. The clusters and the themes in each period segment are as follows:

- *Pre-2014*: ("testing," "data integration," "queuing model," "healthcare modeling," and "bed management," "breast cancer," and more).
- *2014-2016:* ("formal verification," "work sampling," "diagnoses," "car sharing," and more.
- *2016-2018:* ("layout design," "artificial intelligence," "automation," "object-oriented modeling," and more).
- *2018-2020:* ("core manufacturing simulation," "manufacturing planning," "fast-moving consumer goods," "traffic congestion," and more).
- *2020-2022:* ("supply chain planning," "capacity analysis," "virtual reality," "crowd management," "underground mining," "artificial neural network").
- *2022-2024:* ("mass vaccination," "capacity analysis," "load sharing," "ambulance deployment," and more).
- *post-2024:* ("management science," "sales operations planning," "decomposition," "artificial intelligence," "additive manufacturing," "collaborative networks," and more).

The themes dynamics (pre-2014 to 2022-2024 and beyond highlight the trends and evolution in the application of DES to solve diverse problems in several domains in the different period segments identified above. For example, in pre-2014 (2010-2014), themes such as "testing" point to model testing as one of the stages in the DES processes. Also, "data integration" represents one of the tasks during model experimentation where real-life or imaginary data can be integrated with a simulation model to investigate a "what-if" scenario. Other terms such as "healthcare modeling," "breast cancer," and "bed management" point to the application of DES to evaluate cancer diagnoses, hospital bed allocation, and utilization scenarios. Figure 4 shows the complete themes under this category.

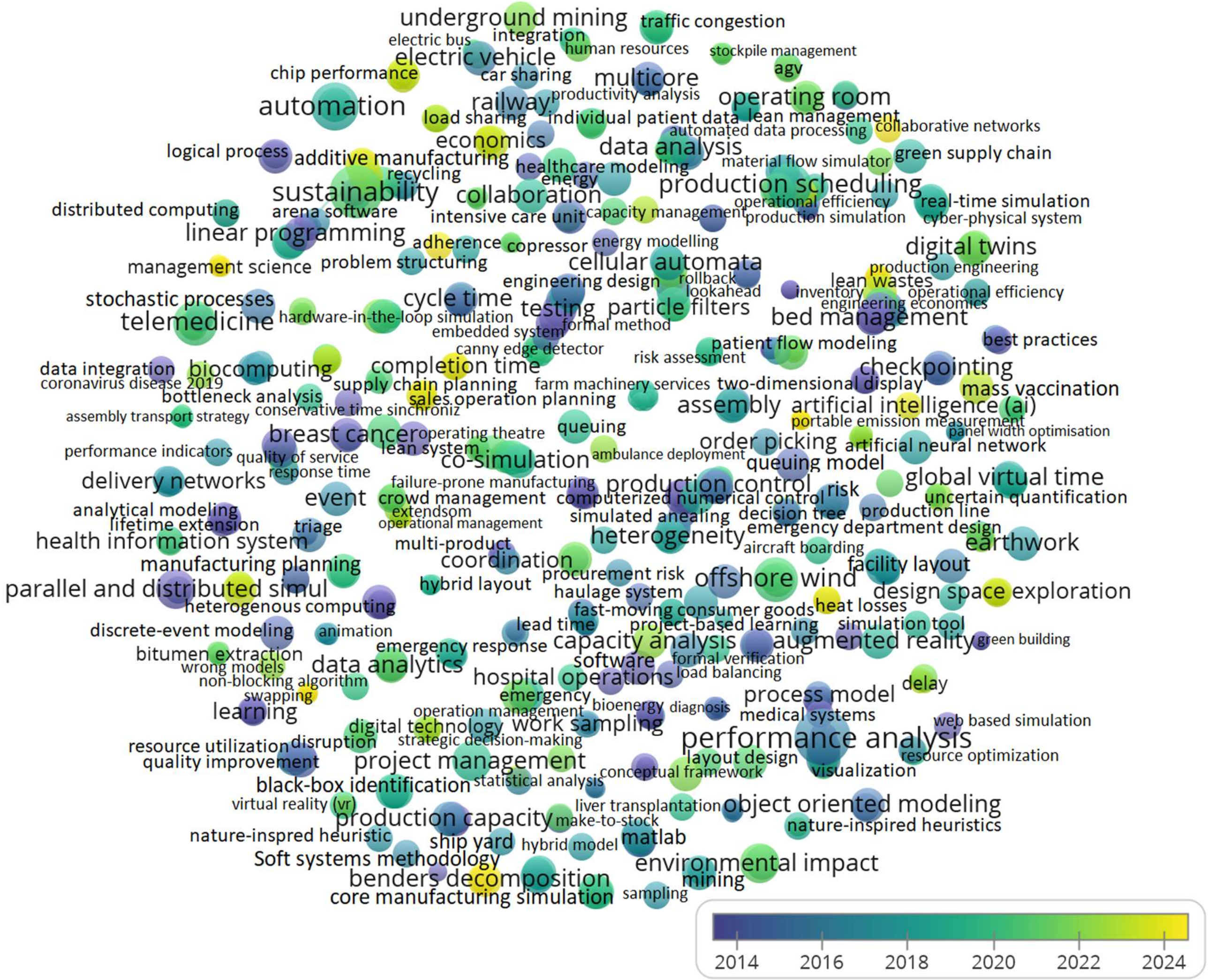

Figure 5. Visualization of the thematic structure involving the least frequent author keywords of four or less occurrences (f < 4).

The period segments 2022-2024, and post-2024 identify themes such as "mass vaccination," "capacity analysis," and "ambulance deployment," pointing to the use of DES to model and simulate critical health issues around COVID-19 vaccination. The current period (2024 and post beyond) mirrors technological trends using DS methods, such as machine learning methodology and "generative adversarial networks" [65]. The entire period (2010-2024) provides the evolutionary trends that highlight the continuous application of DES in research and problem-solving in real-life issues over the years.

### 4.5 Intellectual Structure of DES Research

This Section evaluates the intellectual structure of DES research based on citation analysis in the era of "simulation everywhere" [24] and focuses on the impact of citations on documents and sources. The results help to address the third research objective (RO3). The preliminary results (Table 2) show an average of 9.5 citations per document (Table 2) and a total of 19,731 from 2010-2024 based on SCOPUS bibliographic data. The Google Scholar (scholar.google.com) citation count can be more than the records on SCOPUS.

The citation structure analysis using the R-based Bibliometrix application (Table 4) shows that about 80% of the articles earned at least one or more citations, with the four highest citation impacts occurring in 2013 (95%), 2018 (93%), 2016 (89%), and 2015 (88%). Also, 35% of the current year publications with zero citable years received at least one citation. Table 4 presents the complete citation structure of the publications.

Table 4. The citation structure of research publications on DES.

| Year | >=300 | >=200 | >=100 | >=50 | >=30 | >=10 | >=1 | NC | TP | % Cited |
|---|---|---|---|---|---|---|---|---|---|---|
| 2010 | 0 | 0 | 4 | 7 | 6 | 18 | 56 | 20 | 111 | 82% |
| 2011 | 0 | 1 | 0 | 11 | 4 | 22 | 56 | 26 | 120 | 78% |
| 2012 | 1 | 0 | 3 | 5 | 8 | 33 | 60 | 27 | 137 | 80% |
| 2013 | 0 | 0 | 0 | 3 | 10 | 38 | 67 | 6 | 124 | 95% |
| 2014 | 0 | 0 | 2 | 5 | 8 | 32 | 56 | 17 | 120 | 86% |
| 2015 | 0 | 0 | 0 | 4 | 8 | 38 | 69 | 16 | 135 | 88% |
| 2016 | 0 | 0 | 1 | 5 | 12 | 35 | 89 | 18 | 160 | 89% |
| 2017 | 0 | 0 | 0 | 4 | 6 | 41 | 79 | 25 | 155 | 84% |
| 2018 | 0 | 0 | 1 | 3 | 6 | 38 | 68 | 9 | 125 | 93% |
| 2019 | 0 | 0 | 0 | 2 | 6 | 31 | 83 | 26 | 148 | 82% |
| 2020 | 0 | 0 | 0 | 6 | 0 | 37 | 94 | 26 | 163 | 84% |
| 2021 | 0 | 0 | 0 | 3 | 4 | 27 | 86 | 27 | 147 | 82% |
| 2022 | 0 | 0 | 0 | 0 | 2 | 25 | 87 | 30 | 144 | 79% |
| 2023 | 0 | 0 | 0 | 0 | 0 | 11 | 76 | 54 | 141 | 62% |
| 2024 | 0 | 0 | 0 | 0 | 0 | 1 | 51 | 95 | 147 | 35% |
| Total Pubs | 1 | 1 | 11 | 58 | 80 | 427 | 1077 | 422 | 2077 | |

*NC and TP denote: no citation (publications that did not earn any citations as of December 31, 2024), TP: total publications per year, respectively.*

#### 4.5.1 Most Cited Documents

Table 5 presents the top ten most cited publications. The results show that the article examining "DES and system dynamics in the logistics and supply chain context" [58] recorded the highest citation count. Table 4 presents the complete list of titles and themes and the citation. Fifty percent of the ten most cited articles or conference papers had near maximum citable years. This result implies that citable years accounted for the high citation impacts, among other factors, such as the article title or the sources in which it appears. The most cited article appeared in some of the most popular sources ("Decision Support Systems" and "Journal of Operational Research Society."

Table 5. Ten most cited documents on the use of DES to solve COVID-19 problems

| Rank | Paper | Focus | Journal | TC | AC P/ Year | Citable Years |
|---|---|---|---|---|---|---|
| 1 | [58] | DES and system dynamics in the logistics and supply chain context. | Decision Support Systems | 347 | 24.79 | 14 |
| 2 | [59] | Applications of simulation within the healthcare context. | Journal of the Operational Research Society | 211 | 17.58 | 12 |
| 3 | [60] | The application of DES in a health care setting. | Medical Decision Making | 187 | 13.36 | 14 |
| 4 | [61] | SimLean: Utilizing simulation in the implementation of lean in healthcare. | European Journal of Operational Research | 184 | 14.15 | 13 |
| 5 | [62] | A review of DES application in healthcare. | BMC Health Services Research | 171 | 21.38 | 8 |
| 6 | [63] | Current and future trends of DES and virtual reality use in industry. | IEEE Transaction on Human-Machine System | 169 | 16.9 | 10 |
| 7 | [22] | Combining system dynamics and discrete-event simulation in healthcare. | Proc. Winter Simulation Conf. | 129 | 8.6 | 15 |
| 8 | [64] | Comparing agent-based simulation with DES in modeling emergency behaviors. | Proc. Winter Simulation Conf. | 115 | 7.7 | 15 |
| 9 | [40] | DES method in construction engineering and management. | Journal of Construction Engineering Management | 110 | 7.33 | 15 |
| 10 | [23] | Comparing DES and systems dynamics application in healthcare. | European Journal of Operational Research. | 108 | 9 | 12 |

*TC: Total citations based on SCOPUS bibliographic data; AC P/Year: Average citation per year.*

#### 4.5.2 Eminent Sources

The 2,077 documents analyzed in this study were published in 947 sources, giving an average of 2.2% publications per source. The analysis of eminent sources shows that the top twenty journals and proceedings published 23.3% of the documents and earned 7,635 (38.7%) citations out of 19,731. Most of the journals and conference proceedings were core sources in OR, while others are sources in the application domains. Examples are "Proceedings - Winter Simulation Conference," "Simulation," and "Journal of Simulation." These three sources published 205, 29, 23 articles and proceeding papers and earned citation counts of 1667, 304, and 430. "Others are " Simulation Modelling Practice and Theory," "European Journal of Operational Research," and " Journal of the Operational Research Society." Sources in the application domains include "Medical Decision Making," "Health Care Management Science" (healthcare) and "Automation in Construction" (engineering and construction). Table 6 presents the complete list of the topmost eminent sources.

Table 5. The top 20 sources with most eminent sources and citation impacts.

| Rank | Sources | NP | TC | AVTC | Pub_Start_ Yr. |
|---|---|---|---|---|---|
| 1 | Proceedings - Winter Simulation Conference | 205 | 1667 | 8.1 | 2010 |
| 2 | Simulation | 23 | 430 | 18.7 | 2010 |
| 3 | Journal of Simulation | 29 | 304 | 10.5 | 2010 |
| 4 | Simulation Modelling Practice and Theory | 17 | 621 | 36.5 | 2011 |
| 5 | European Journal of Operational Research | 12 | 622 | 51.8 | 2010 |
| 6 | Computers and Industrial Engineering | 11 | 330 | 30.0 | 2014 |
| 7 | Journal of the Operational Research Society | 14 | 386 | 27.6 | 2010 |
| 8 | Medical Decision Making | 16 | 449 | 28.1 | 2010 |
| 9 | Procedia CIRP | 20 | 217 | 10.9 | 2014 |
| 10 | Value In Health | 13 | 429 | 33.0 | 2010 |
| 11 | IFAC-Papers Online | 16 | 166 | 10.4 | 2015 |
| 12 | ACM Transactions on Modeling and Computer Simulation | 19 | 181 | 9.5 | 2011 |
| 13 | Automation in Construction | 9 | 342 | 38.0 | 2012 |
| 14 | BMC Health Services Research | 9 | 317 | 35.2 | 2011 |
| 15 | Health Care Management Science | 7 | 288 | 41.1 | 2011 |
| 16 | International Journal of Advanced Manufacturing Technology | 9 | 155 | 17.2 | 2011 |
| 17 | Lecture Notes in Comp Science; Subseries Lecture Notes in Artificial Intelligence; Lecture Notes in Bioinformatics | 28 | 158 | 5.6 | 2010 |
| 18 | PLOS ONE | 11 | 186 | 16.9 | 2011 |
| 19 | Proceedings of the 2013 Winter Simulation Conference - Simulation: Making Decisions in A Complex World | 12 | 107 | 8.9 | 2013 |
| 20 | Journal of Construction Engineering and Management | 6 | 280 | 46.7 | 2010 |

*NP: No. of Publications; TC: Total citations; AVTC: Average TC Per Document; Pub Yr.: Publication Start Year*

### 4.6 Social Structure of Publications

#### 4.6.1 Countries' Publications and Impact

The final part of the analysis examines the social structure of publications on DES research, which addresses the fourth research objective (RO4). Recently, the simulation community began paying attention to "simulation around the world," which has become a track at conferences, such as the Winter Simulation Conference [65]. The results help to answer Collaboration among countries where the authors are domiciled is another measure of the social structure of publications in science mapping studies [50,51,53].

The results show 1,968 unique authors from 89 countries that publish articles used in this study, with 5,429 total frequencies as some authors published more than one articles. The international collaboration index results show moderate association of 19.03% among the authors' affiliated countries and international co-authorship.

Table 6 shows the USA, the UK, China, Germany, and Italy as the five most dominant countries in literature publications on DES and applications. The Table 6 shows a complete list of the top ten countries. The analysis also highlights the countries where the corresponding authors are domiciled, which follow a similar pattern as literature publications led by USA, the UK, and Italy. The USA published the highest number of articles (200, including 35 as corresponding authorship) through collaborations, the UK came second with 98 documents through global collaborations but had 22 corresponding authorships. The USA and UK also earned the most citations (2,517, 2,315, respectively). Table 6 shows the complete list of the top 10 countries where the authors/co-authors are domiciled.

Table 6. Countries international collaborations in publications in two categories, SCP and MCP and impact.

| Country | Publications | SCP | MCP | TC | ATC(Pub) |
|---|---|---|---|---|---|
| United States | 200 | 165 | 35 | 2517 | 12.59 |
| United Kingdom | 98 | 76 | 22 | 2315 | 23.62 |
| China | 79 | 70 | 9 | 703 | 8.9 |
| Germany | 73 | 60 | 13 | 995 | 13.63 |
| Italy | 62 | 47 | 15 | 693 | 11.18 |
| Canada | 60 | 52 | 8 | 1214 | 20.23 |
| Brazil | 44 | 36 | 8 | 530 | 12.05 |
| Sweden | 36 | 23 | 13 | 470 | 13.06 |
| France | 33 | 23 | 10 | 198 | 6 |
| Australia | 30 | 19 | 11 | 746 | 24.87 |
| India | 29 | 25 | 4 | 118 | 4.07 |
| Korea | 24 | 22 | 2 | 267 | 11.13 |
| Malaysia | 24 | 20 | 4 | 147 | 6.13 |
| Turkey | 24 | 15 | 9 | 285 | 11.88 |
| Indonesia | 22 | 20 | 2 | 57 | 2.59 |

*SCP: Collaboration as non-corresponding author; MCP: Contributions as corresponding author; TC (Total Citations); ATC(Pub): Average total citation per publication*

## 5. Conclusions

This study attempts to advance the body of knowledge on the important discourse about the continuous relevance of discrete-event simulation (DES) and its applications. As an OR method, DES has been an effective optimization technique since the 1950s [1-3] and a decision support system for decades [7,12,28,66]. The empirical evidence in this study shows that DES is alive, well, and kicking.

The results show a continuous expansion of DES as an OR technique, including when researchers and practitioners raised concerns about its future relevance. The research productivity recorded asymmetrical but positive yearly output with an annual average of 2.03%, while the application areas broadened with a sustained relevance as a scientific method for tackling old, new, and emerging issues. For example, DES remained an essential OR tool in tackling COVID-19-related problems [25]. In recent years, DES has remained at the center of digital transformation that has swept across the industrial space, from manufacturing to healthcare and the service industry, hence the notion of simulation everywhere. DES also played a central role in the various aspects of the fourth and fifth industrial revolutions (Industry 4.0, 5.0, respectively). Some notable and vital areas include digital twins, virtual manufacturing, and human-robot interaction [67-70]. Other sectors, such as healthcare/medical applications and the general digital transformation sweeping across the industrial space, from manufacturing to healthcare and other sectors, also have significant DES footprints. The impact analysis also showed significantly high performance, as per the analysis of the third research objective in this study.

The prospects look positive. Further, with the emergence and ongoing adoption and deployment of generative artificial intelligence (GenAI) [71,72], the simulation community must seek ways to integrate GenAI in DES to remain relevant and improve the modeling and simulation processes.

**Author Contributions:** For research articles with several authors, a short paragraph specifying their individual contributions must be provided. The following statements should be used "Conceptualization, I.J.A.; methodology, I.J.A.; software, I.J.A. and G.E.; validation, I.J.A. and G.E.; formal analysis, I.J.A.; investigation, I.J.A.; resources, I.J.A.; and G.E.; data curation, I.J.A.; writing—original draft preparation, I.J.A.; writing—review and editing, I.J..A. and G.E.; visualization, I.J.A. and G.E.; supervision, I.J.A.; project administration, I.J.A. and G.E. All authors have read and agreed to the published version of the manuscript."

**Funding:** This research received no external funding.

**Data Availability Statement:** Proprietary data used; authors are not authorized to share the data.

**Acknowledgments:** We thank the Almighty God for the gift of live, health, and salvation.

**Conflicts of Interest:** None declared.